Article

# Characterization of a modified clinical linear accelerator for ultra-high dose rate beam delivery

Umberto Deut[1,2§]\*, Aurora Camperi[2§], Cristiano Cavicchi[3], Roberto Cirio[1,2], Emanuele Data[1,2], Elisabetta Durisi[1,2], Veronica Ferrero[1,2], Arianna Ferro[1,2], Simona Giordanengo[2], Oscar A. Martì Villarreal[4], Felix Mas Milian[1,2,5], Elisabetta Medina[1,2], Diango M. Montalvan Olivares[1,2], Franco Mostardi[1,2], Valeria Monti[1,2], Roberto Sacchi[1,2], Edoardo Salmeri[3], Anna Vignati[1,2†]

[1] Università degli Studi di Torino, Dipartimento di Fisica, Torino, Italy;
[2] National Institute of Nuclear Physics (INFN), Sezione di Torino, Torino, Italy;
[3] Elekta S.p.A., Agrate Brianza (MB), Italy;
[4] Fondazione Bruno Kessler (FBK), Center for Sensors and Devices, Trento, Italy;
[5] Universidade Estadual de Santa Cruz, Department of Exact and Technological Sciences, Ilheus, Brazil;
\* Correspondence: umberto.deut@unito.it;
§ These authors contributed equally to the work and share first authorship
† Last author

**Abstract:** Irradiations at Ultra High Dose Rate (UHDR) regimes, exceeding 40 Gy/s in single fractions lasting less than 200 ms, have shown an equivalent antitumor effect compared to conventional radiotherapy with reduced harm to normal tissues. This work details the hardware and software modifications implemented to deliver 10 MeV UHDR electron beams with a Linear Accelerator Elekta SL 18 MV and the beam characteristics obtained. GafChromic EBT XD films and an Advanced Markus chamber were used for the dosimetry characterization, while a silicon sensor assessed the machine's beam pulses stability and repeatability. Dose per pulse, average dose rate and instantaneous dose rate in the pulse were evaluated for four experimental settings, varying the source-to-surface distance and the beam collimation, i.e. with and without the use of a cylindrical applicator. Results showed dose per pulse from 0.6 Gy to a few tens of Gy and average dose rate up to 300 Gy/s. The obtained results demonstrate the possibility to perform in-vitro radiobiology experiments and test of new technologies for beam monitoring and dosimetry at the upgraded LINAC, thus contributing to the electron UHDR research field.

## 1. Introduction

The research on Ultra High Dose Rate Radiotherapy (UHDR RT) is experiencing significant growth, prompting the scientific community to investigate the reasons behind the normal tissue sparing effect (FLASH effect) and to take the necessary steps towards its potential translation into the clinical context [1-6].

Although the beam characteristics able to trigger the FLASH effect are far to be fully understood, there is consensus on the relevance of some irradiation parameters' thresholds. UHDR RT requires instantaneous dose rates around $10^5$ Gy/s and at least 40 Gy/s as average dose rates [1,2], with single irradiation fractions lasting less than 200 ms. [7]

The availability of machines capable of delivering UHDR beams is crucial for every aspect of the research: from radiobiology experiments [2, 5] to the development of dedicated Treatment Planning Systems (TPS) [8, 9]. For this reason, interest in the modification of clinical LINACs has increased over the last decade [10-12].

This work presents the reversible modification of a LINAC Elekta SL 18 MV towards the delivery of 10 MeV electron UHDR beam and its dosimetry characterization in different irradiation settings, i.e. varying the source-to-surface distance (SSD) and with the optional use of a cylindrical polymethyl methacrylate (PMMA) applicator.

Dosimetry measurements in UHDR modality were performed with GafChromic films EBT XD and compared to the results obtained in conventional regime with both films and an Advanced Markus (AM) ionization chamber. Thin silicon sensors, placed at the exit of the accelerator head, were used to monitor the output beam pulses and evaluate their temporal uniformity within the pulse as well as the pulse-by-pulse stability.

This upgrade of the LINAC Elekta SL 18 MV facility provides the researchers a valuable tool for UHDR studies, allowing testing different beam monitor technologies and performing radiobiological experiments. Additionally, the use of perforated templates positioned at the applicator's end allow studying the characteristics of Spatially Fractionated Radiation Therapy (SFRT), by positioning perforated templates along the beamline [13].

## 2. Materials and Methods

The LINAC Elekta SL 18 MV was installed in 2016 at the Physics Department of the University of Turin (UNITO) and is entirely dedicated to research. This machine can generate both X-rays (up to 18 MV) and electrons (4 – 18 MeV), delivering the beam in 2 µs pulses at a Pulse Repetition Frequency (PRF) between 6 and 400 Hz. The accelerator head is equipped with multi-leaf collimators (MLC) and diaphragms which precisely shape the beam, allowing radiation fields up to 40 x 40 cm² at the isocenter position. It also contains a monitoring system consisting of two independent ionization chambers (ICs), one acting as the primary reference and the second as a backup.

The 10 MeV electron beam can be delivered both in the standard LINAC clinical configuration (conventional mode) and with the high dose rate LINAC upgrade described in section 2.1 (UHDR mode).

The beam pulse characterization with a silicon sensor and dose measurements, respectively described in section 2.2 e 2.3, were performed for four irradiation settings, each one featuring different SSDs and collimation of the beam (Fig. 1):

*(i)*      at the isocenter position (SSD = 100 cm);

*(ii)*      at the isocenter position with a PMMA cylindrical applicator of 5 cm inner diameter;

*(iii)*      at the crosshair foil (SSD = 52.9 cm), at the exit of the LINAC head;

*(iii)*      in the wedge tray (SSD = 18.6 cm), inside the LINAC head.

For all the settings the irradiation field size was fixed at 10 x 10 cm² at the isocenter position *(i)* resulting in smaller field sizes for the configurations closer to the source.

The crosshair foil position *(iii)* is the closest point to the source where standard dosimetry procedures can still be applied, while the wedge tray position *(iv)* can only be evaluated with in-air measurements.

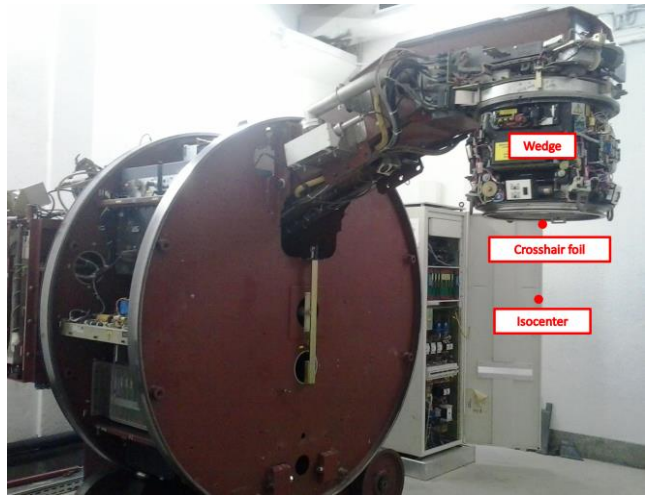

**Figure 1.** The LINAC Elekta SL 18 MV with three (out of the four) irradiation settings available: isocenter (*i*), crosshair foil (*iii*) and wedge (*iv*). The setting relying on the PMMA applicator (*ii*) is not shown.

### 2.1. Clinical Accelerator Upgrade

The upgrade of the LINAC aiming to generate a UHDR electron beam was supervised by Elekta technicians, who implemented both hardware and software modifications to the machine, following the work of Lempart et al. [10].

In the control console, the default parameters of the 10 MV photon mode are adjusted to deliver an electron beam, preventing the X-ray target from interfering with the beam path. All the filters, including the wedge and the shutter foil, are removed, leaving a free spot of approximately 100 cm³ on the wedge tray between the ICs and the light mirror. This spot corresponds to the irradiation setting *iv*, listed above.

The electron gun filament current is increased from the conventional 5.6 A to 7.3 A. The bending magnet intensity is tuned to optimize the beam transport and maximize the accelerated beam current under the new conditions. The power drawn from the magnetron is increased to approximately 6 kW, which remains below the specified maximum power of 7 kW.

Each time the LINAC is started, a standard warm-up procedure is performed by delivering about 1000 Monitor Units (MU) of a 15 MV X-ray beam and about 1000 MU of the 10 MeV electron clinical beam to reach the optimal working conditions. Once the regime is switched to UHDR modality, the parameters related to the electron gun current, gun aim, gun standby and tuner control are manually adjusted to the optimal values. The parameters generally experience small variations during the optimization, and the operation usually requires a few beam deliveries for prior adjustments. During LINAC operation, the water temperature is maintained between 26 °C and 28 °C, a range in which the best performance can be achieved.

Specific interlocks must be overridden to operate the LINAC in UHDR conditions [10, 12]. To avoid the interlock from the internal ICs, a custom attenuator circuit provided by Elekta was connected to the LINAC. This circuit attenuates the current signal arriving from the ICs, preventing the interlock from being activated.

The attenuator circuit makes the ICs reading unreliable, so they cannot be used to monitor the dose delivered. Therefore, a pulse counter circuit (PCC) was developed to control the LINAC delivery in UHDR mode by counting the number of beam pulses [10, 11].

The circuit converts the current signal generated in an unbiased silicon diode sensor into voltage signals using a transimpedance amplifier. The voltage signals are then filtered through a Sallen-Key filter and amplified to produce an acceptable input signal for a Schmitt Trigger. The resulting 5 V amplitude square pulses are counted using an Arduino

NANO board. The PCC is connected to the LINAC allowing the trigger pulse to reach the thyratron only until the desired amount of beam pulses is delivered.

All measurements reported in this study are referred to the High Power (HP) mode of the RF injection cycle of the magnetron, where two charging cycles occur before the thyratron is triggered by the Pulse-Forming Network (PFN). HP mode is needed for reaching the highest possible dose rates. However, a Low Power (LP) mode is also available, resulting in dose rate values intermediate between conventional and UHDR. The availability of three LINAC modalities (conventional, UHDR LP and UHDR HP) can be exploited to vary the delivered dose, while maintaining the same experimental setup.

All the reported modifications do not compromise the operation of the LINAC in the conventional mode, and few minutes are required to switch between the irradiation modalities.

### 2.2. Pulse characterization with silicon sensors

The temporal structure and the consistency of the output pulses after the LINAC modification were measured with a silicon sensor previously tested for response linearity up to doses of 10 Gy/pulse at the SIT ElectronFlash accelerator (9 MeV UHDR electron beam) at Centro Pisano FLASH Radiotherapy (CPFR, Pisa) [14]. The sensor was manufactured by the Fondazione Bruno Kessler (FBK) and consists of an epitaxial substrate grown on a low-resistivity silicon layer. This device is a square of 4.5 mm edge with an active thickness of 45 μm and a support layer of 570 μm. It is segmented in eight pads with different active areas (0.03 mm² - 2.3 mm²) isolated by guard rings. For the measurements, the device was mounted on a high voltage (HV) distribution PCB (Fig. 2a), designed to polarize the sensor from the back via a copper layer with a window behind the sensor, so that only a 2 mm thick layer of FR4 resin is behind the sensor. Bonding wires (Al 1% Si) with a diameter of 25 μm, allowing a maximum rated current of 500 mA, were used to connect the 2 mm² pad to an output channel of the board and to ground the guard ring.

The multipad sensor was reverse biased at 200 V and the stability of the HV supply was monitored during the acquisitions to guarantee the stability of the applied voltage. The sensor was manually positioned at the center of the field at the crosshair foil position (*iii*) with an uncertainty of 5 mm in the transverse plane.

The output channel was connected via a coaxial cable to a Keysight Infiniium S-Series oscilloscope (2.5 GHz, 20 GSa/s, Model: DSOS254A), to read and store the output pulse signals delivered by the LINAC (Fig. 2b). The produced charge per pulse was determined by dividing the signal area by the input impedance of the oscilloscope (50 Ω). The signal processing involved only the voltage offset removal from the signal calculated as the average of the first off-pulse 10000 samples.

The stability and reproducibility of the delivered pulses are then studied with repeated measurements of the charge per pulse generated under consecutive pulses and on different days.

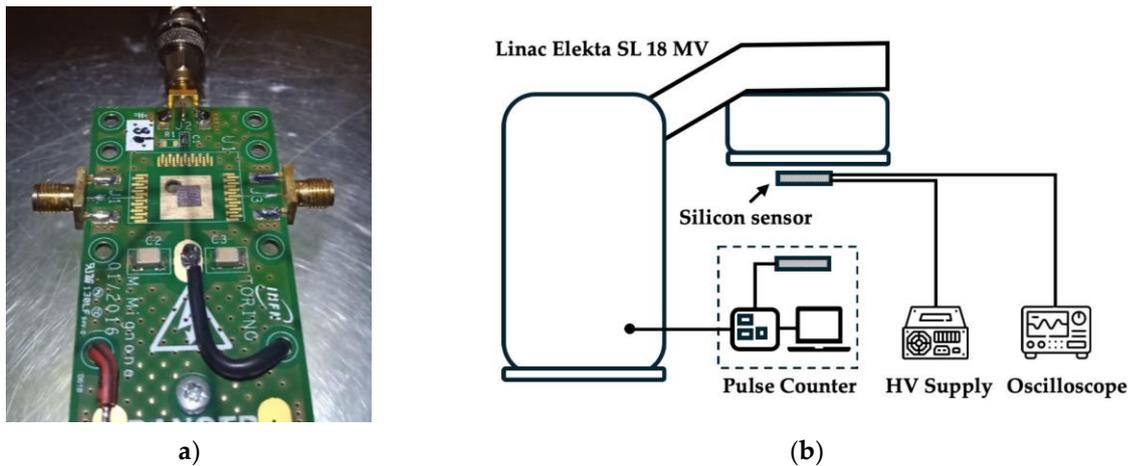

**a)**                                                        **(b)**

**Figure 2.** (**a**) PCB with bonded sensor in the centre; (**b**) Schematic of the setup with the silicon sensor at the crosshair foil position (*iii*).

## 2.3. Dosimetry measurements

The beam energy verification after the upgrade was performed by measuring the percentage depth dose (PDD) curves in both conventional and UHDR modalities. In the first case, a complete PDD curve was obtained using the Advanced Markus (AM) chamber (Type 34045, PTW, Freiburg, Germany) and a 1D scanner water tank (Sun Nuclear Corporation, Melbourne, USA) (Fig. 3a), following the TRS-398 guidelines of the International Atomic Energy Agency (IAEA) [15].

In the case of UHDR irradiations, PDD curves were collected by placing EBT XD Gaf-Chromic films (Ashland, Bridgewater, USA) between PMMA slabs to avoid distortions due to the large charge recombination observed in ICs at large dose rates [16, 17].

PDD curves were measured with the water phantom and the PMMA slabs at the isocenter position both with and without the applicator (irradiation settings *ii* and *i*, respectively). A holder secured the applicator at the crosshair foil position, with its exit side aligned with the surface of the water tank or the PMMA (Fig. 3b). The detector was always placed transversely at the center of the field.

From these curves, it was possible to calculate dosimetry parameters such as the practical range ($R_P$) and the depths where the absorbed dose was 50% ($R_{50}$) and 80% ($R_{80}$) of the maximum. Based on $R_{50}$, the reference depth ($z_{ref}$), the position at which the evaluation of the absorbed dose should be performed, was calculated as recommended in [15]:

$$z_{ref} = 0.6R_{50} - 0.1 \tag{1}$$

EBT XD GafChromic films were used to quantify the beam output in terms of total dose, average Dose Per Pulse (DPP), instantaneous dose rate in the pulse and the average dose rate. The instantaneous dose rate in the pulse was obtained from the average DPP divided by the pulse duration $\tau$ (2 μs), while the average dose rate was calculated from the number of delivered pulses (N) at a fixed PRF of 100 Hz, as reported by Cetnar et al. [12].

$$Average\ Dose\ Rate = \frac{total\ dose}{\frac{N-1}{PRF} + \tau} \tag{2}$$

For these measurements, EBT XD films were placed in the PMMA phantom with a buildup thickness of 18 mm. This thickness was selected based on the $z_{ref}$ values determined from the PDDs analysis and considering the correction factors for transforming the water thickness into the equivalent PMMA thickness. In the wedge tray position, due to space restrictions, the films were irradiated without buildup.

Furthermore, GafChromic films were also used to evaluate the beam spatial profiles in both irradiation modalities. When using the films, the total number of pulses N was changed according to the setting, to avoid film saturation.

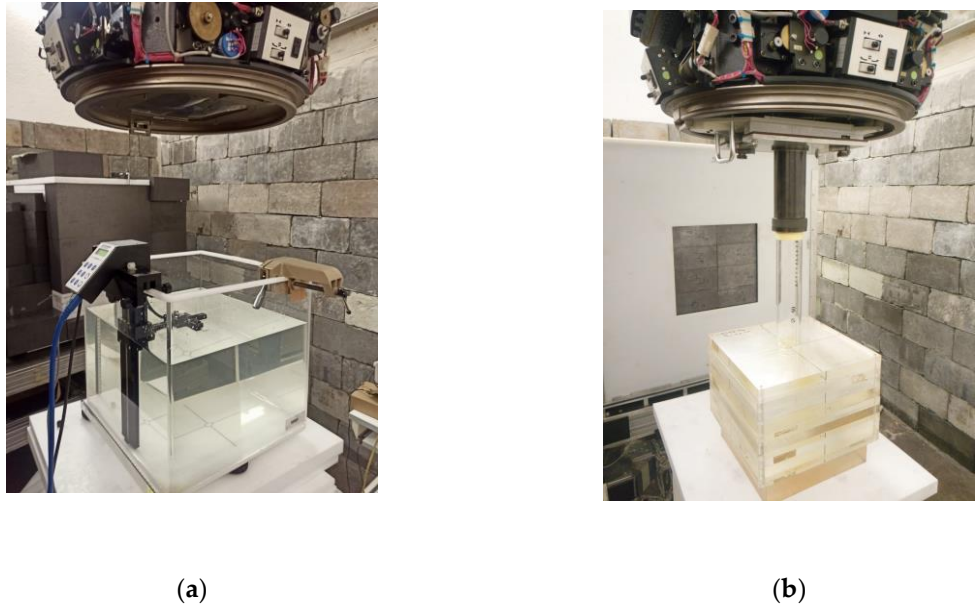

(**a**)                                                                     (**b**)

**Figure 3.** (**a**) 1d water phantom positioned at the isocenter (*i*); (**b**) setup with the applicator (*ii*) and the PMMA phantom

The absolute dose of the irradiated films was evaluated in terms of net Optical Density (OD), as established in the manufacturer guidelines and protocols [18, 19]. For this purpose, the films were scanned with a flatbed color scanner (Model: Epson Expression 12000XL) used in transmission mode, acquiring 48-bit RGB with a resolution of 100 dpi. The films were scanned twice: before the irradiation to save the background image and 24 hours after irradiation to determine the absorbed dose. The analysis of the images has been executed with a program developed in Matlab, where the net OD was calculated from the pixel-by-pixel difference between the OD of the irradiated and the background images. The net ODs were calculated over a Region of Interest (ROI) of the same size to the active area of the AM chamber ($0.20 \ cm^2$). The conversion from net OD to dose was executed with two different calibration curves: the first one relying on the 10 MeV conventional electron beam in the 0.5 - 10 Gy range, and the second one on the 9 MeV UHDR electron beam delivered at the CPFR (Pisa, Italy) for the range 0.5 – 40 Gy. Below the 10 Gy threshold, the observed difference between the calibration curves was less than 5%, which corresponds to the standard uncertainty associated with the dose obtained with a GafChromic film [11]. For doses higher than 10 Gy, only the calibration relying on the CPFR irradiations was used. For further reliability, a calibration curve was also determined by scanning the films one week after irradiation. The difference between the curves referring to 24 hours and one week is 2.5% (Fig. 4).

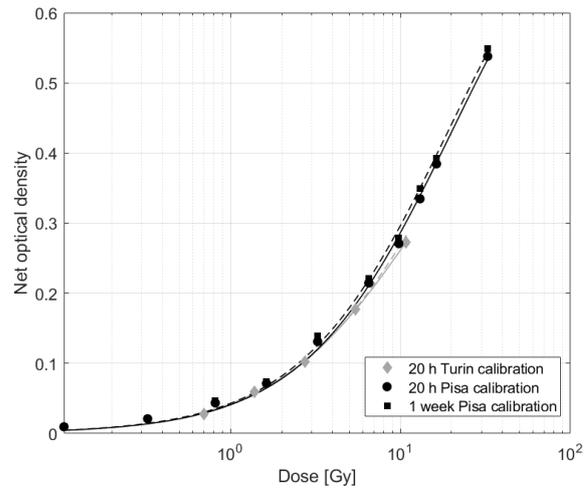

**Figure 4.** Different EBT XD calibrations with best fit curves.

All the calibration curves were fitted with the following rational function

$$netOD = a + \frac{b}{Dose - c} \qquad (3)$$

where a, b and c are free parameters.

Finally, a study of the Output Factor (OF) was conducted in both irradiation modalities at the isocenter position (*i*), varying the beam field from 3 x 3 cm² to 30 x 30 cm².

## 3. Results

### 3.1. Beam Pulse Characterization

The monitoring of the beam with the silicon sensor allows a qualitative and quantitative analysis of the beam pulses. A comparison of the voltage pulses acquired with the oscilloscope at the different irradiation settings positions is shown in figure 5.

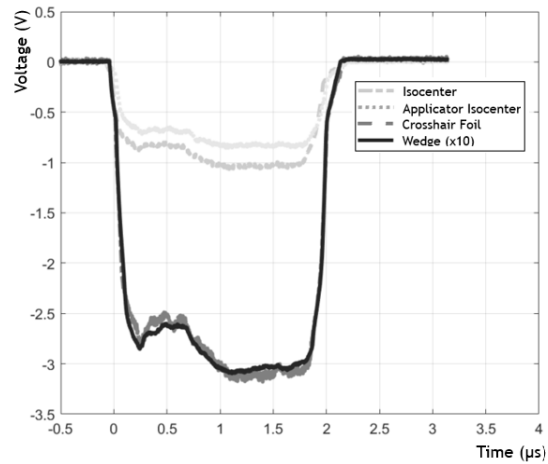

**Figure 5.** Output voltage pulse generated by the 10 MeV UDHR electron beam from the LINAC on a silicon sensor (60 um thickness) recorded by the oscilloscope with the different irradiation settings. The wedge signal has been divided by a factor of 10 for ease of comparison.

The figure demonstrates the capability of increasing, through the optimization procedure, the charge per pulse while maintaining the 2 µs pulse shape and duration mostly unchanged. Figure 6 shows the charge collected per pulse in the UHDR modality for qualifying the performance of the accelerator in this modality.

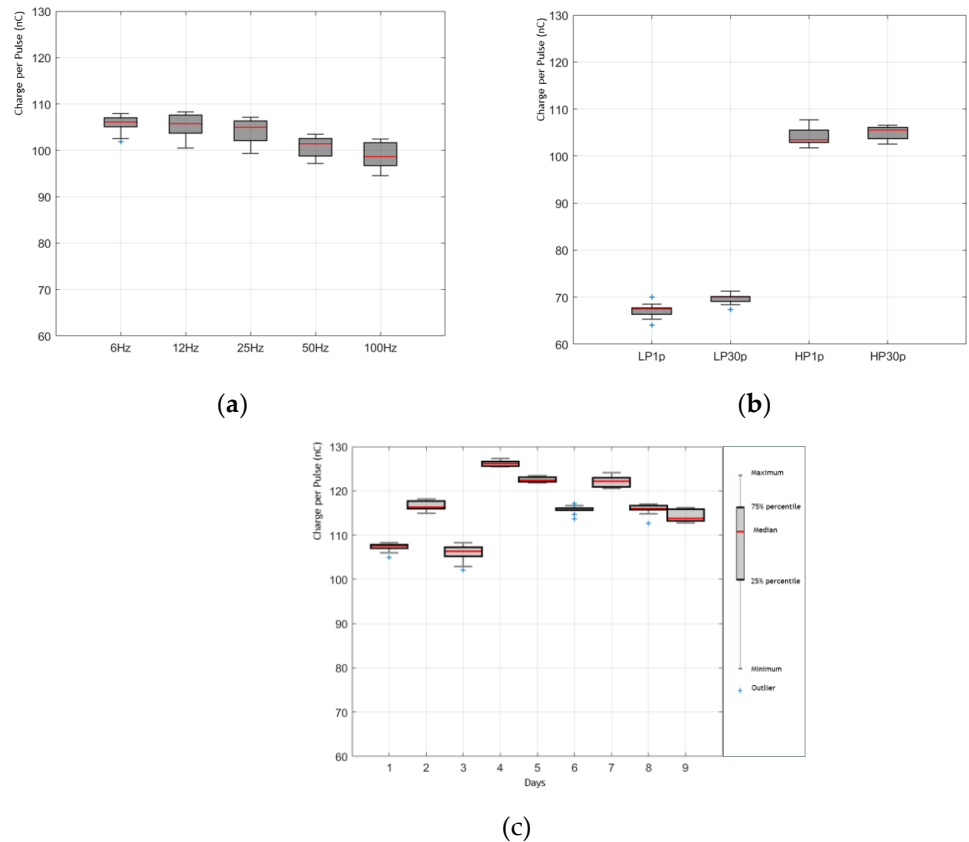

**Figure 6.** (**a**) Collected charge per pulse at the crosshair foil position for different PRF. Each boxplot represents a total of 20 pulses delivered in HP mode; (**b**) comparison of the charge per pulse at the crosshair foil position in LP and HP modes considering single pulse irradiations repeated 10 times (LP1p and HP1p) and 30 consecutive pulses (LP30p and HP30p; (**c**) repeatability of the charge per pulse at the crosshair foil position acquired in different days in HP mode.

In Figure 6a, the collected charge per pulse at the crosshair foil position irradiating 20 consecutive pulses is presented as a function of the PRF. A 7% decrease of the median charge value is observed between the lowest and highest PRFs, while, at the same PRF, the maximum pulse-by-pulse deviation from the median value is 5%. The dose per pulse in ten single-pulse deliveries and in a delivery with 30 consecutive pulses are compared in Figure 6b for the two RF injection cycle modes (LP and HP), showing a maximum difference of the 3.7%. The LP mode exhibits a better pulse-by-pulse stability at the expense of a lower charge per pulse, which appears to be 35% lower compared to the HP mode. Indeed, consecutive beam deliveries in LP mode result in a variation of the charge of 2% while in HP the maximum deviation from the median is 3.8%. The reproducibility of the pulses in HP mode is presented in Figure 6c, where the charge per pulse at the crosshair measured in identical irradiations of 30 pulses with PRF of 100Hz was repeated in nine different days within three months. The deviation from the median value is less than 8% for the 90% of the measurements performed.

### 3.2. Dosimetry

The UHDR PDD curves obtained with GafChromic films show good agreement with the curves obtained with both the AM chamber and films for the conventional 10 MeV electron beam (Fig. 7). The maximum difference between $R_{50}$ values is 3.6% and only the initial points (near to the surface) show an underdose that could be caused by some air gaps between the PMMA slabs or by the proximity to the film cut [20]. The dosimetry parameters mentioned in section 2.3 were calculated and are shown in Tab. 1.

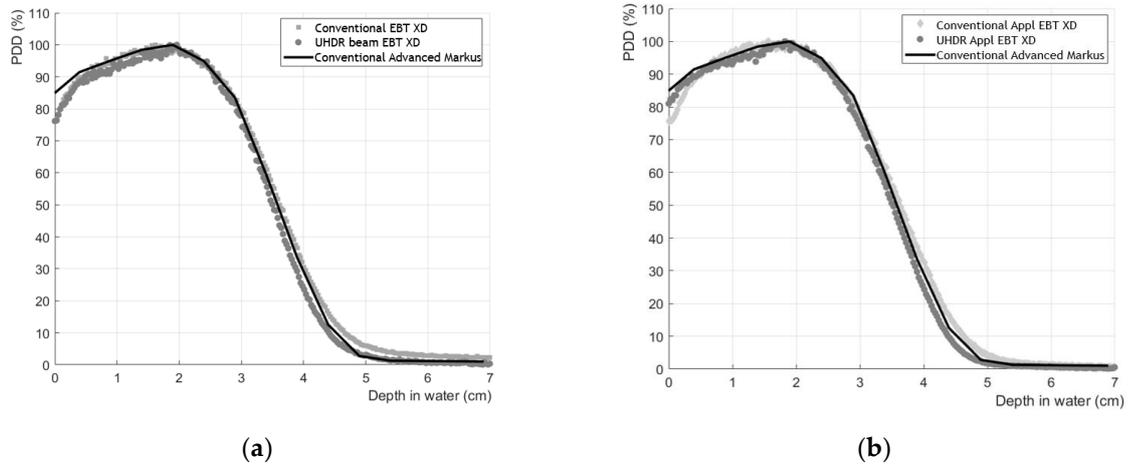

**Figure 7.** Comparison between PDD curves measured with the Advanced Markus and GafChromic films EBT XD in conventional and UHDR regime: (**a**) without applicator (**b**) with applicator.

**Table 1.** Comparison of range parameters in water in conventional and UHDR mode.

| Mode | Setup | $R_{80}$ (cm) | $R_{50}$ (cm) | $R_P$ (cm) | $z_{ref}$ (cm) |
|---|---|---|---|---|---|
| Conventional | Advanced Markus | 2.95 ± 0.10 | 3.61 ± 0.11 | 4.75 ± 0.18 | 2.07 ± 0.07 |
| | EBT XD | 2.94 ± 0.07 | 3.65 ± 0.05 | 4.69 ± 0.26 | 2.09 ± 0.03 |
| | EBT XD with applicator | 2.97 ± 0.05 | 3.67 ± 0.02 | 4.72 ± 0.24 | 2.10 ± 0.01 |
| UHDR | EBT XD | 2.88 ± 0.10 | 3.54 ± 0.10 | 4.55 ± 0.28 | 2.03 ± 0.06 |
| | EBT XD with applicator | 2.89 ± 0.04 | 3.54 ± 0.02 | 4.55 ± 0.19 | 2.03 ± 0.01 |

The dosimetry values estimated with different detectors are comparable with each other and confirm that the modification of the LINAC did not affect the energy distribution of the beam, even when the applicator is inserted. The uncertainties of the $R_{80}$ and $R_{50}$ values are at most 3%, while the $R_P$ values are characterized by larger uncertainties of 5% since they are calculated as the intersection between two lines. The $R_{50}$ values reported in Table 1 show are 7% smaller than the expected value for 10 MeV electrons [21], indicating that the most probable beam energy is slightly less than 10 MeV.

From the absorbed dose, measured for the four irradiation settings at the reference depth $z_{ref}$, it was possible to extract the DPP, the average dose rate (Eq. 2), and the instantaneous dose rate in the pulse. Results are reported in Table 2

As expected from previous studies [22], the comparison between conventional and UHDR modalities indicates an increase of the three dose quantities, for all the irradiation settings, of at least three orders of magnitude. At the wedge tray position (*iv*), in UHDR regime, only one pulse was delivered to avoid exceeding the dose range of EBT XD films (40 Gy); thus, it was not possible to calculate the average dose rate for multiple pulses as in the other settings.

**Table 2.** Dosimetry results

| | | Dose Per Pulse (Gy) | | Average Dose Rate (Gy/s) | | Instantaneous Dose Rate (Gy/s) | |
|---|---|---|---|---|---|---|---|
| | mode | conv | UHDR | conv | UHDR | conv | UHDR |
| Setting | Isocenter | $(1.60 \pm 0.08) \cdot 10^{-4}$ | $0.63 \pm 0.04$ | $(1.60 \pm 0.08) \cdot 10^{-2}$ | $83.6 \pm 4.2$ | $(7.99 \pm 0.40) \cdot 10^{1}$ | $(3.60 \pm 0.18) \cdot 10^{5}$ |
| | Applicator | $(2.52 \pm 0.13) \cdot 10^{-4}$ | $0.81 \pm 0.04$ | $(2.52 \pm 0.13) \cdot 10^{-2}$ | $89.5 \pm 4.5$ | $(1.26 \pm 0.06) \cdot 10^{2}$ | $(4.03 \pm 0.20) \cdot 10^{5}$ |
| | Crosshair foil | $(6.76 \pm 0.34) \cdot 10^{-4}$ | $2.22 \pm 0.11$ | $(6.76 \pm 0.34) \cdot 10^{-2}$ | $309 \pm 16$ | $(3.38 \pm 0.17) \cdot 10^{2}$ | $(12.3 \pm 0.6) \cdot 10^{5}$ |
| | Wedge* | $(0.92 \pm 0.05) \cdot 10^{-2}$ | $27.2 \pm 1.4$ | $(92.3 \pm 4.6) \cdot 10^{-2}$ | n.d.** | $(4.61 \pm 0.23) \cdot 10^{3}$ | $(136 \pm 7) \cdot 10^{5}$ |

\* Dose in air without build up
\*\* n.d.: not determined

The use of the PMMA applicator results in an increase in the DPP at the isocenter of 57% and 28% in conventional and UHDR modes, respectively.

The beam profiles obtained from the scanned images are shown in Figure 8.

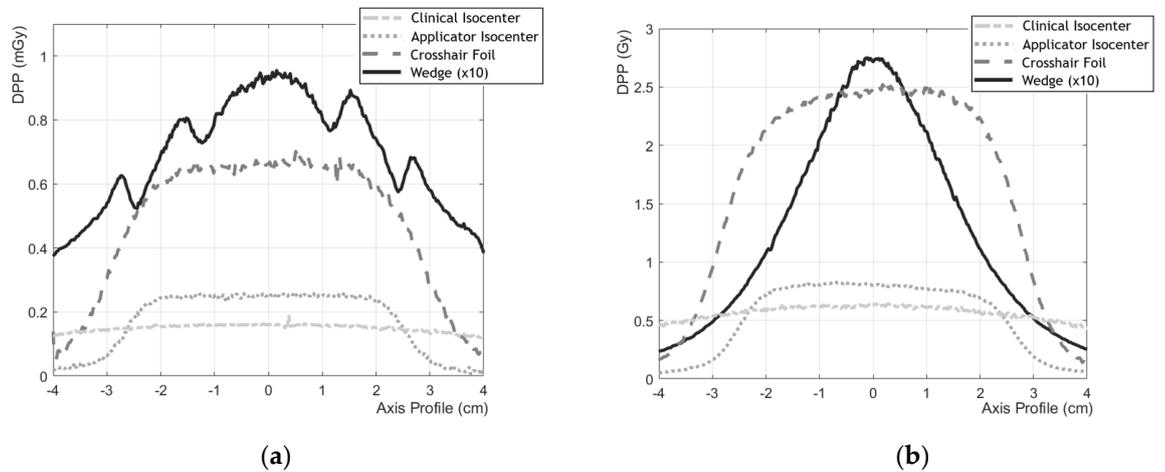

**Figure 8.** Dose profiles obtained from GafChromic films for the conventional (**a**) and UHDR mode (**b**). In both figures the dose profiles at the wedge tray (setting *iv*) were divided by a factor of 10 for ease of comparison.

In conventional modality it's possible to note few bumps for the wedge setting profile. This can be attributed mainly to the geometry of the primary filters and of the secondary scattering foils, which have two sloped sections corresponding to specific scattering angles [10].

The study on the output factor in both irradiation modalities is reported in figure 9. In the conventional mode, there is a good agreement between the AM chamber and the GafChromic films. The analysis of the UHDR mode data show a similar trend with the conventional ones. For fields larger than 10x10 cm², UHDR values are systematically smaller as reported by Dal Bello et al. [10], but all the differences are inside the error bars.

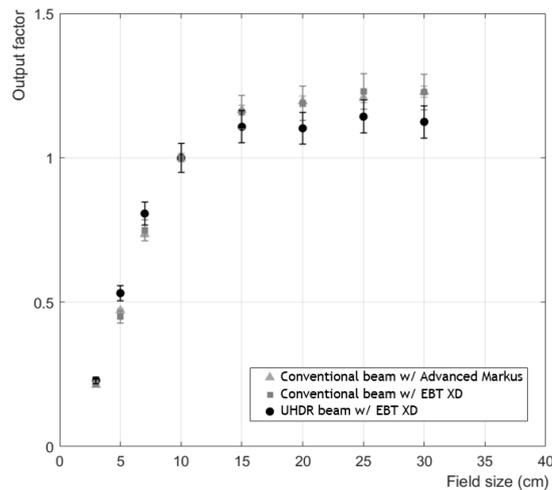

**Figure 9.** Output factor measured with the Advanced Markus and GafChromic films EBT XD in conventional and UHDR mode for different square field sizes.

## 4. Discussion

In this study, a LINAC Elekta SL 18 MV was upgraded to deliver a 10 MeV electron beam in UHDR conditions, and a characterization of the beam was performed. The main modifications consisted of removing the primary and secondary scattering foils, increasing the gun current and magnetron power and adjusting some parameters related to beam transport. Moreover, a custom PCC was developed to count the pulses delivered by the LINAC and stop the irradiation after the requested number of pulses, due to the unreliability of the ICs chamber after the addition of an attenuator.

A silicon diode sensor, whose response has already been proven to be linear with the DPP [14], was used to verify the stability of the pulses within the same beam delivery and their repeatability across different beam deliveries and days. A variation of 3.7% in the charge per pulse was measured between single and consecutive pulse deliveries. Different irradiations within the same day showed good reproducibility with a maximum charge variation from the median of 2% for the LP mode and 4% for the HP mode. It is important to note that the UHDR pulse stability verification of the LINAC was performed considering all pulses, including the first one, which, as shown, can be up to 4% smaller than the stable pulses. For this reason, the percentage variation shown in this study was slightly higher than that of Konradsson et al. [23], who modified a similar Elekta LINAC and reported deviations of less than 3% when the first pulse was excluded from the analysis. Since it is known that standby periods may impact on the performance of the LINAC [23], the reproducibility was assessed in different days over a period of three months. The deviation of the charge per pulse measured by the silicon diode was less than 8% from the median value and is deemed acceptable for in-vitro radiobiology studies.

The beam energy distribution was not affected by the LINAC modifications, as demonstrated by the PDD curves and the related parameters. The DPPs and the dose rates achieved in this study are comparable to those obtained for other modified LINACs [10, 12], providing from a minimum 0.6 Gy/pulse ($3.6 \cdot 10^5$ Gy/s instantaneous dose rate) at the isocenter position (*i*) to a maximum of 27 Gy/pulse ($136 \cdot 10^5$ Gy/s instantaneous dose rate) at the wedge tray position (*iv*). Average dose rate values over 300 Gy/s were reached at the crosshair foil position (*iii*) with a PRF of 100 Hz. These values allow planning both radiobiology experiments and test of instrumentation for comparison in conventional and UHDR regimes, although the limited space available in the wedge irradiation setting needs to be carefully considered to design proper arrangements.

Several improvements of the reached performances will be investigated in the next months. Considering that the maximum PRF reachable for this LINAC is 400 Hz and that the magnetron power could be increased to 7 kW, further optimization of the delivery parameters will be studied aiming at reaching higher DPP and average dose rates. Exploiting the capability of silicon sensors to resolve single delivered pulses, a correlation of single pulses measurements and delivered dose will contribute to the study of a dose-based beam monitoring strategy, moving from the one based on the number of pulses.

The results obtained in this work confirm PMMA applicators as a good option to increase the dose at the isocenter. This could be exploited in conjunction with perforated templates to disentangle the contribution to tissue-sparing effects due to UHDR and spatial fractionated beams [24]. The spatial dose distribution and the possible collimation techniques could be studied under the same settings proposed in this study. The dosimetry characteristics such as the dose peak's full width half maximum and peak to valley dose ratio could be characterized at different dose rates. The different energies available could help contributing to the characterization of c with a clinical energy spectrum.

Finally, the possibility to deliver 18 MeV electron beams will be carefully considered to explore UHDR irradiations at higher energies [25].

## 5. Conclusions

The LINAC Elekta SL 18 MV has been successfully upgraded to deliver 10 MeV UHDR electron beams. This modification is completely reversible and switching between the conventional modality and the UHDR modality takes only a few minutes. The LINAC is now capable of reaching a maximum of 2.2 Gy/pulse and over $10^5$ Gy/s of instantaneous dose rate at the crosshair foil position in a pulse lasting 2 µs. A silicon sensor device assesses the stability and repeatability of the pulses across different beam deliveries and days.

In the future, the LINAC could be exploited as a facility for testing beam monitoring and dosimetry devices, as well as for radiobiological experiments to further investigate the FLASH effect. The use of applicators with different geometry and various perforated templates could enable the study of the effects of SFRT and its combination with UHDR irradiation.

**Author Contributions:** Conceptualization, U.D., A.C., C.C., R.C., E.Da., E.Du., V.F., A.F., S.G., O.A.M.V., F.M.M., E.M., D.M.M.O., F.M., V.M., R.S., E.S. and A.V.; methodology, U.D., A.C., C.C., R.C., E.Da., E.Du., V.F., A.F., S.G., O.A.M.V., F.M.M., E.M., D.M.M.O., F.M., V.M., R.S., E.S. and A.V.; software, U.D., A.C., C.C., E.Da., E.Du., A.F., O.A.M.V., F.M.M., E.M., D.M.M.O., F.M., V.M. and E.S.; validation, U.D., A.C., C.C., R.C., E.Da., E.Du., V.F., A.F., S.G., F.M.M., E.M., D.M.M.O., F.M., V.M., R.S., E.S. and A.V.; formal analysis, U.D., A.C., E.M. and D.M.M.O; investigation, U.D., A.C., C.C., R.C., E.D., E.D., V.F., A.F., S.G., O.A.M.V., F.M.M., E.M., D.M.M.O., F.M., V.M., R.S., E.S. and A.V.; resources, U.D., A.C., C.C., E.Da., E.Du., E.M., D.M.M.O., V.M., and E.S.; data curation, U. D. and A.C.; writing – original draft preparation, U.D., A.C. and D.M.M.O.; writing – review and editing, U.D., A.C., C.C., E.Du., S.G., E.M., D.M.M.O., V.M., R.S., E.S. and A.V.; visualization, U.D., A.C., C.C., R.C., E.Da., E.Du., V.F., A.F., S.G., O.A.M.V., F.M.M., E.M., D.M.M.O., F.M., V.M., R.S., E.S. and A.V.; supervision, E. Du., S.G., V.M., R.S. and A.V.; project and administration, R.C., S.G., R.S. and A.V.; funding acquisition, R.C., S.G., R.S. and A.V.

**Funding:** The research was supported by the INFN CSN5 project "FRIDA" and by VIGA S1921 EX-POST 21 01 (Compagnia di San Paolo). The installation of the LINAC at the Physics Department was financed by Compagnia di San Paolo grant "OPEN ACCES LABS" (2015), Fondazione CRT grant n.2015.AI1430.U1925 and INFN.

**Data Availability Statement:** The raw data supporting the conclusions of this article will be made available by the authors, without undue reservation.

**Acknowledgments:** We thank INFN CSN5 funded project "eXFlu" for the collaboration and "Fondazione Pisa" for funding CPFR with the grant "prog. n. 134/2021". The authors are also grateful to the Elekta S.p.A. for the technical support in the LINAC upgrade and maintenance.

**Conflicts of Interest:** Cristiano Cavicchi and Edoardo Salmeri are employees of Elekta S.p.A., Agrate Brianza (MB), Italy. The remaining authors declare that the research was conducted in the absence of any commercial or financial relationships that could be construed as a potential conflict of interest.

# References

1. Favaudon V, Caplier L, Monceau V, Pouzoulet F, Sayarath M, Fouillade C, et al. Ultrahigh dose-rate FLASH irradiation increases the differential response between normal and tumor tissue in mice. *Sci Transl Med*. (**2014**) 6:1–10. doi:10.1126/scitranslmed.3008973.

2. Montay-Gruel P, Petersson K, Jaccard M, Boivin G, Germond J.F, Petit B, Doenlen R, Favaudon V, Bochud F, Bailat C, et al. Irradiation in a Flash: Unique Sparing of Memory in Mice after Whole Brain Irradiation with Dose Rates above 100 Gy/s. *Radiotherapy and Oncology,* **2017**, *124*, 365–369, doi:10.1016/j.radonc.2017.05.003.

3. Vozenin MC, De Fornel P, Petersson K, Favaudon V, Jaccard M, Germond JF, Petit B, Burki M, Ferrand G, Patin D, Bouchaab H, Ozsahin M, Bochud F, Bailat C, Devauchelle P, Bourhis J. The Advantage of FLASH Radiotherapy Confirmed in Mini-pig and Cat-cancer Patients. Clin Cancer Res. **2019** Jan 1;25(1):35-42. doi: 10.1158/1078-0432.CCR-17-3375. Epub 2018 Jun 6. PMID: 29875213.

4. Hageman, E., Che, P.-P., Dahele, M., Slotman, B. J. & Sminia, P. *Radiobiological Aspects of FLASH Radiotherapy. Biomolecules*, https://doi.org/10.3390/biom12101376 (**2022**).

5. Di Martino F., Barca P., Barone S., Bortoli E., Borgheresi R., De Stefano S., Di Francesco M., Faillace L., Giuliano L., Grasso L., Linsalata S., Marfisi D., Migliorati M., Pacitti M., Palumbo L., Felici G. FLASH Radiotherapy With Electrons: Issues Related to the Production, Monitoring, and Dosimetric Characterization of the Beam. *Frontiers in Physics* **2020**, vol. 8, doi: 10.3389/fphy.2020.570697

6. Vignati A, Giordanengo S, Fausti F, Martì Villarreal OA, Mas Milian F, Mazza G, Shakarami Z, Cirio R, Monaco V and Sacchi R (**2020**) Beam Monitors for Tomorrow: The Challenges of Electron and Photon FLASH RT. Front. Phys. 8:375. doi: 10.3389/fphy.2020.00375.

7. Giannini, N.; Gadducci, G.; Fuentes, T.; Gonnelli, A.; Di Martino, F.; Puccini, P.; Naso, M.; Pasqualetti, F.; Capaccioli, S.; Paiar, F. Electron FLASH Radiotherapy in Vivo Studies. A Systematic Review. *Front. Oncol.* **2024**, *14*, 1373453, doi:10.3389/fonc.2024.1373453.

8. Rahman, M.; Trigilio, A.; Franciosini, G.; Moeckli, R.; Zhang, R.; Böhlen, T.T. FLASH Radiotherapy Treatment Planning and Models for Electron Beams. *Radiotherapy and Oncology* **2022**, *175*, 210–221, doi:10.1016/j.radonc.2022.08.009.

9. Lempart, M.; Blad, B.; Adrian, G.; Bäck, S.; Knöös, T.; Ceberg, C.; Petersson, K. Modifying a Clinical Linear Accelerator for Delivery of Ultra-High Dose Rate Irradiation. Radiotherapy and Oncology 2019, 139, 40–45, doi:10.1016/j.radonc.2019.01.031.

10. Dal Bello, R.; Von Der Grün, J.; Fabiano, S.; Rudolf, T.; Saltybaeva, N.; Stark, L.S.; Ahmed, M.; Bathula, M.; Kucuker Dogan, S.; McNeur, J.; et al. Enabling Ultra-High Dose Rate Electron Beams at a Clinical Linear Accelerator for Isocentric Treatments. *Radiotherapy and Oncology* **2023**, *187*, 109822, doi:10.1016/j.radonc.2023.109822.

11. Cetnar AJ, Jain S, Gupta N, Chakravarti A. Technical note: Commissioning of a linear accelerator producing ultra-high dose rate electrons. *Med Phys*. **2024**; 51:1415–1420. https://doi.org/10.1002/mp.16925

12. Schüler, E.; Trovati, S.; King, G.; Lartey, F.; Rafat, M.; Villegas, M.; Praxel, A.J.; Loo, B.W.; Maxim, P.G. Experimental Platform for Ultra-High Dose Rate FLASH Irradiation of Small Animals Using a Clinical Linear Accelerator. *International Journal of Radiation Oncology\*Biology\*Physics* **2017**, 97, 195–203, doi:10.1016/j.ijrobp.2016.09.018.

13. Yan W, Khan MK, Wu X, Simone CB 2nd, Fan J, Gressen E, Zhang X, Limoli CL, Bahig H, Tubin S, Mourad WF. Spatially fractionated radiation therapy: History, present and the future. *Clin Transl Radiat Oncol*. **2019** Oct 22;20:30-38. doi: 10.1016/j.ctro.2019.10.004. PMID: 31768424; PMCID: PMC6872856.

14. Medina E, Ferro A, Abujami M, Camperi A, Centis Vignali M, Data E, Del Sarto D, Deut U, Di Martino F, Fadavi Mazinani M, Ferrero M, Ferrero V, Giordanengo S, Martì Villarreal OA, Hosseini MA, Mas Milian F, Masturzo L, Montalvan Olivares DM, Montefiori M, Paternoster G, Pensavalle JH, Sola V, Cirio R, Sacchi R and Vignati A, First experimental validation of silicon-based sensors for monitoring ultra-high dose rate electron beams. *Front. Phys.* **2024**. 12:1258832. doi: 10.3389/fphy.2024.1258832

15. INTERNATIONAL ATOMIC ENERGY AGENCY *Absorbed Dose Determination in External Beam Radiotherapy*; Technical Reports Series; Rev. 1.; INTERNATIONAL ATOMIC ENERGY AGENCY, **2024**; ISBN 978-92-0-146022-6.

16. Subiel, A.; Moskvin, V.; Welsh, G.H.; Cipiccia, S.; Reboredo, D.; DesRosiers, C.; Jaroszynski, D.A. Challenges of Dosimetry of Ultra-Short Pulsed Very High Energy Electron Beams. *Physica Medica* **2017**, *42*, 327–331, doi:10.1016/j.ejmp.2017.04.029.

17. McManus, M., Romano, F., Lee, N.D. et al. The challenge of ionisation chamber dosimetry in ultra-short pulsed high dose-rate Very High Energy Electron beams. *Sci Rep* 10, 9089 (**2020**). https://doi.org/10.1038/s41598-020-65819-y.

18. Niroomand-Rad, A., Blackwell, C.R., Coursey, B.M., Gall, K.P., Galvin, J.M., McLaughlin, W.L., Meigooni, A.S., Nath, R., Rodgers, J.E. and Soares, C.G. (**1998**), Radiochromic film dosimetry: Recommendations of AAPM Radiation Therapy Committee Task Group 55. Med. Phys., 25: 2093-2115. https://doi.org/10.1118/1.598407

19. Devic, S., Seuntjens, J., Sham, E., Podgorsak, E. B., Schmidtlein, C. R., Kirov, A. S., & Soares, C. G. (**2005**). Precise radiochromic film dosimetry using a flat-bed document scanner. *Medical physics*, 32(7Part1), 2245-2253.

20. Costa F, et al., Assessment of clinically relevant dose distributions in pelvic IOERT using Gafchromic EBT3 films, *Physica Medica* (**2015**), http://dx.doi.org/10.1016/j.ejmp.2015.05.013

21. Strydom W, Parker W, Olivares M. Chapter 8 electron beams: Physical and clinical aspects. In: Radiation Oncology Physics: A Handbook for Teachers and Students. Montreal, McGill University: IAEA; 2006.p. 1-46

22. Ashraf MR, Rahman M, Zhang R, Williams BB, Gladstone DJ, Pogue BW and Bruza P (**2020**) Dosimetry for FLASH Radiotherapy: A Review of Tools and the Role of Radioluminescence and Cherenkov Emission. *Front. Phys.* 8:328. doi: 10.3389/fphy.2020.00328.

23. Konradsson Elise, Wahlqvist Pontus, Thoft Andreas, Blad Börje, Bäck Sven, Ceberg Crister, Petersson Kristoffer. Beam control system and output fine-tuning for safe and precise delivery of FLASH radiotherapy at a clinical linear accelerator. *Frontiers in Oncology*, 14, **2024**, https://www.frontiersin.org/journals/oncology/articles/10.3389/fonc.2024.1342488, doi:10.3389/fonc.2024.1342488. 2234-943X

24. Pensavalle Jake Harold, Romano Francesco, Celentano Mariagrazia, Sarto Damiano Del, Felici Giuseppe, Franciosini Gaia, Masturzo Luigi, Milluzzo Giuliana, Patera Vincenzo, Prezado Yolanda, Di Martino Fabio. Realization and dosimetric characterization of a mini-beam/flash electron beam. *Frontiers in Physics*. Vol. 11. **2023**. https://www.frontiersin.org/journals/physics/articles/10.3389/fphy.2023.1269495. doi:10.3389/fphy.2023.1269495. 2296-424XL.

25. Faillace, D. Alesini, G. Bisogni, F. Bosco, M. Carillo, P. Cirrone, G. Cuttone, D. De Arcangelis, A. De Gregorio, F. Di Martino, V. Favaudon, L. Ficcadenti, D. Francescone, G. Franciosini, A. Gallo, S. Heinrich, M. Migliorati, A. Mostacci, L. Palumbo, V. Patera, A. Patriarca, J. Pensavalle, F. Perondi, R. Remetti, A. Sarti, B. Spataro, G. Torrisi, A. Vannozzi, L. Giuliano, Perspectives in linear accelerator for FLASH VHEE: Study of a compact C-band system, *Physica Medica*, Volume 104, **2022**, Pages 149-159, ISSN 1120-1797, https://doi.org/10.1016/j.ejmp.2022.10.018.